\documentclass[10pt]{JHEP3}
\usepackage{graphicx}
\usepackage[centertags]{amsmath}
\usepackage{array,multirow}
\usepackage{amssymb}
\usepackage{graphicx}
\usepackage{color}

\usepackage{epsfig}
\def\cf{{\mathcal F}}
\def\cd{{\mathcal D}}
\def\cfhat{\hat{\cf}}
\def\cfmus{{\mathcal F}_{\mu_1\ldots\mu_s}}
\def\cfhmus{{\cfhat}_{\mu_1\ldots\mu_s}}
\def\gsym[#1]{g_{(\mu_1\mu_2}{{#1}_{\mu_3\ldots\mu_s)}}}
\def\gsymup[#1]{g^{(\mu_1\mu_2}{{#1}^{\mu_3\ldots\mu_s)}}}
\def\phit{\tilde{\phi}}

\def\cz{{\mathcal Z}}

\def\l{\left}
\def\r{\right}

\def\m{\mu}

\def\a{\alpha}

\def\>{\rangle}
\def\<{\langle}
\def\ZZ{\mathcal{Z}}

\newcommand{\pb}{\bar p} 
\newcommand{\rb}{\bar r} 
\newcommand{\WW}{\mathcal{W}} 
\preprint{HRI/ST/1203\\NIKHEF2012-014} 
\title{Partition Functions for Higher-Spin theories in $AdS$} 
\author{Rajesh Kumar Gupta$^{a}$\footnote{rgupta AT nikhef DOT nl} and Shailesh Lal$^{b}$\footnote{shailesh AT hri DOT res DOT in} \\
$^a$NIKHEF theory group, Science Park 105,\\
$\;$1098 XG Amsterdam,\\
$\;$The Netherlands\\
$^b$Harish-Chandra Research Institute, \\
$\;$Chhatnag Road,\\
$\;$Jhusi, India 211019\\}
\abstract{We calculate the one-loop partition function for a massless arbitrary-spin field on quotients of a general dimensional $AdS$ background using the results of arXiv:1103.3627. We use these results to compute the one-loop partition function for a Vasiliev theory in $AdS_5$. An interesting form of the answer, suggestive of a vacuum character of an enhanced symmetry algebra is obtained. We also observe a close connection between the partition function for this Vasiliev theory and the $d$-dimensional MacMahon function.}
\begin{document}
\section{Introduction}
The holography of higher-spin theories has been a topic of some sustained exploration \cite{Klebanov:2002ja}--\cite{Vasiliev:2012vf}. This interest has essentially been fuelled by two motivations. Firstly, standard $AdS$/CFT arguments lead us to believe that the twist-two sector of a (free, planar) CFT$_d$---such as $\mathcal N=4$ SYM---would be described in the $AdS_{d+1}$ bulk by a consistent classical theory with an infinite tower of arbitrary-spin particles. It is natural to expect that this bulk theory would be a Vasiliev theory. See \cite{HaggiMani:2000ru} for details and \cite{Bekaert:2005vh} for a review of Vasiliev theories in general dimensions. Since a free CFT is in some sense a natural starting point for organising the boundary theory into a string theory (see \cite{Gopakumar:2003ns,Gopakumar:2004qb,Gopakumar:2005fx} for a systematic approach) probing these questions would be helpful in order to learn about the general mechanics of gauge-string duality. Equally intriguingly, there is the additional possibility that even when the boundary theory is not free, higher-spin symmetries are still present in a higgsed phase \cite{Beisert:2004di}. This, if true, would afford us insight into the full set of gauge invariances of string theory. Secondly, in the low-dimensional examples of $AdS_3$ and $AdS_4$, these theories are expected to have CFT duals in their own right \cite{Klebanov:2002ja,Gaberdiel:2010pz,Minwalla:2011ma,Aharony:2011jz,Giombi:2011kc}. Typically, the best known and studied examples of $AdS$/CFT are supersymmetric \cite{Maldacena:1997re}, because supersymmetry allows us the freedom to reliably compute and extrapolate results on both sides of the duality. However, there are good reasons to believe that the general phenomena of gauge-string duality and holography are not tied to supersymmetry. A natural allied question is to ask what role \textit{does} supersymmetry play in $AdS$/CFT.  In this regard, the holography of higher-spin theories is interesting because many of these dualities are non-supersymmetric \cite{Klebanov:2002ja,Gaberdiel:2010pz,Chang:2011mz} and offer us an opportunity to concretely explore $AdS$/CFT away from supersymmetry. 

Studies of the asymptotic symmetries of higher-spin theories in $AdS_3$ have recently led to remarkable progress in their holography, including the formulation of an explicit duality between Vasiliev theories in $AdS_3$ and $\mathcal{W}_N$ minimal models \cite{Gaberdiel:2010pz}. It was shown in \cite{Henneaux:2010xg,Campoleoni:2010zq} that the asymptotic symmetry algebra of (a class of) higher-spin theories comprises of two copies of a $\WW$ algebra. This was checked in \cite{Gaberdiel:2010ar} by computing the one-loop partition function of the theory, to obtain a vacuum character of left- and right-moving $\WW$ algebras. The results of \cite{Gaberdiel:2010ar} were in fact a key ingredient in formulating the duality of \cite{Gaberdiel:2010pz}. The analysis carried out in \cite{Gaberdiel:2010ar} drew heavily on the heat kernel methods of \cite{David:2009xg}. Essentially, the message to take away from these computations is that the second-quantised partition function contains information about additional symmetries which act on \textit{multiparticle} states of the theory.

While a Brown-Henneaux like analysis for Vasiliev theories in dimensions higher than three is prohibitive, the one-loop partition function may be computed readily. The essential ingredients are again the quadratic action for higher-spin fields, which is well known (see \cite{Campoleoni:2010zq} for a recent review), and the determinants of the Laplacian, which have been computed in \cite{Gopakumar:2011qs}.

In this paper, we shall carry out precisely such a computation for a Vasiliev theory in general-dimensional $AdS$ spacetimes. In particular, we shall evaluate this partition function on a thermal quotient of $AdS$. As mentioned above, for the one-loop partition function of a theory the details of the interactions are not important. The only information that enters is the quadratic action and the spectrum. We shall choose the spectrum of all spins $s=0,1,2,\ldots \infty$ appearing once \footnote{As is well known, there is a also minimal Vasiliev theory with only even spins. We consider the theory containing all spins in its spectrum. On the boundary CFT, this corresponds to considering conserved currents built out of complex scalars. Also see \cite{Sezgin:2001zs,Sezgin:2001yf} for representations of the higher-spin algebra in $AdS_5$.}. We shall obtain a form for the partition function which is suggestive of a vacuum character of a symmetry group. We also express the partition function in a form \eqref{macmahon} which is closely related to the $d$-dimensional MacMahon function.

A brief overview of this paper is as follows. In section \ref{partitionsection} we compute the partition function of a massless\footnote{The notion of masslessness for a tensor field is slightly ambiguous on a curved manifold such as $AdS$, by `massless', we mean that the action has a gauge freedom, which we specify in \eqref{gaugetrs}.} spin-$s$ field in $AdS_D$ by the heat kernel method.\footnote{We evaluate these determinants explicitly for odd-dimensional $AdS$ spacetimes. For a very general class of tensors, which includes STT tensors, the analysis for even dimensions greater than two is entirely analogous and we do not repeat it here. See \cite{Gopakumar:2011qs} and references therein.} We show that the degrees of freedom that contribute to the one-loop partition function are encoded in symmetric, transverse, traceless (STT) tensors of rank $s$, and $s-1$. These determinants have been evaluated in \cite{Gopakumar:2011qs} We obtain an answer consistent with the the partition function of the spin-$s$ conserved current in CFT$_{D-1}$. As we outline later (see the discussion just above section \ref{blindsection}), while this analysis goes through for general dimensional $AdS$ spacetimes, though the expressions for $AdS_5$ are perhaps the nicest, and we concentrate on this case in the main body of the paper. 

For evaluating these determinants to arrive at concrete expressions for the partition function, we begin with the relatively simpler case of thermal $AdS_5$, \textit{i.e.} turning off all other chemical potentials, in section \ref{blindsection}. We observe simplifications due to which the partition function of the theory arranges itself into a form reminisctive of the MacMahon function, and also write down a nested, vacuum character like form for this partition function. Expressions for the partition function of the higher-spin theories in $AdS_4$ and $AdS_7$ are available in Appendix \ref{otherdim}.

In section \ref{refinedsection} we generalise to the case of non-zero chemical potentials for the $SO(4)$ Cartans of the $AdS_5$ isometry group. We find that the essential simplifications in section \ref{blindsection} remain, due to which it is still possible to write the answer in terms of a (shifted) two-dimensional MacMahon function. We also obtain a nested form for the answer, suggestive of a vacuum character interpretation. We then discuss our results in section \ref{conclusions}. Some calculational details are available in the Appendices.

\section{The Partition Function for a Higher-Spin Field in $AdS$}
\label{partitionsection}
We begin with computing the partition function of a massless spin-$s$ field in an arbitrary dimensional (Euclidean) $AdS$ spacetime. Our starting point will be the functional integral
\begin{equation}
\label{pathint}
Z^{\left(s\right)}=\frac{1}{\text{Vol(gauge group)}}\int \left[D\phi_{\left(s\right)}\right]e^{-S\left[\phi_{\left(s\right)}\right]},
\end{equation}
which we shall compute in the one-loop approximation. Hence, only the spectrum of the theory and the quadratic terms in the action are of relevance to us. The quadratic action for higher-spin fields has already been worked out in an arbitrary dimensional $AdS$ spacetime \cite{{Fronsdal:1978rb,Fronsdal:1978vb}} (also see the related work \cite{Buchbinder:2001bs,Sagnotti:2003qa}\footnote{We thank M. Tsulaia for bringing these to our attention.}). We shall briefly recollect the main elements of the story.

Our starting point is a quadratic action for a symmetric rank-$s$ tensor field which is invariant under the gauge transformation
\begin{equation}
\label{gaugetrs}
\phi_{\mu_1\ldots \m_s}\mapsto \phi_{\m_1\ldots \m_s}+\nabla_{(\m_1}\xi_{\m_2\ldots \m_s)}.
\end{equation}
For a consistent description, it turns out that the fields $\phi$ necessarily satisfy  a double-tracelessness constraint which is trivial for spins 3 and lower.
\begin{equation} 
\phi_{\m_1\ldots \m_{s-4}\nu\rho}\,^{\nu\rho}=0.
\end{equation}
The gauge transformation parameter $\xi$ then satisfies a tracelessness constraint
\begin{equation}
\xi_{\m_1\ldots \m_{s-3}\nu}\,^{\nu}=0.
\end{equation}
Note that this constraint is non-trivial---and is imposed---even for the spin-3, which has no double tracelessess constraint. 
With these conditions, a quadratic action for the spin-$s$ field in $AdS_D$, invariant under the gauge transformation \eqref{gaugetrs} may be written down.
\begin{equation}
\label{action}
S\l[\phi_{\l(s\r)}\r]=\int\,d^Dx\sqrt{g} \phi^{\mu_1\ldots\mu_s}\l(\cfhat_{\mu_1,\ldots,\mu_s}-\frac{1}{2}{\gsym[\cfhat]}_\lambda\,^\lambda\r),
\end{equation}
where
\begin{equation}
\label{fhat}
\cfhmus=\cfmus-\frac{s^2+\l(D-6\r)s-2\l(D-3\r)}{\ell^2}\phi_{\mu_1\ldots\mu_s}-\frac{2}{\ell^2}{\gsym[\phi]}_\lambda\,^\lambda,
\end{equation}
and
\begin{equation}
\label{f}
\cfmus=\Delta\phi_{\mu_1\ldots\mu_s}-\nabla_{(\mu_1}\nabla^{\lambda}\phi_{\mu_2\ldots\mu_s)\lambda}+\frac{1}{2}\nabla_{(\mu_1}\nabla_{\mu_2}{\phi_{\mu_3\ldots\mu_s)\lambda}}^\lambda.
\end{equation}
We now evaluate the functional integral \eqref{pathint} with the action \eqref{action} using the method adopted in \cite{{Gaberdiel:2010xv,Gaberdiel:2010ar,Bagchi:2011td}} which we now summarise. Since this action has a gauge invariance associated with it, we need to gauge fix. Typically this involves the introduction of Faddeev-Popov ghosts. In this case however, there is a natural choice of integration variables for the functional integral. We use the decomposition
\begin{equation}
\label{decomposition}
\phi_{\mu_1\ldots\mu_s}=\phi^{TT}_{\mu_1\ldots\mu_s}+\gsym[\phit]+\nabla_{(\mu_1}\xi_{\mu_2\ldots\mu_s)},
\end{equation}
where $\phit$ and $\xi$ are respectively symmetric traceless rank $s-2$ and $s-1$ tensors. We note that the last term is just the gauge transformation of the field $\phi$ under which the action \ref{action} is invariant. This is the choice of integration variables employed in \cite{Gaberdiel:2010xv} for the spin-2 field, and subsequently generalised in \cite{Gaberdiel:2010ar} to higher-spin fields in $AdS_3$. 
The measure for the functional integral changes under this change of variables.
\begin{equation}
[D\phi_{(s)}]=\cz^{(s)}_{gh}[D\phi^{TT}_{(s)}][D{\phit}_{(s-2)}]\l[D\xi_{(s-1)}\r],
\end{equation}
where $\cz^{(s)}_{gh}$ is the ghost determinant that arises from making the change of variables. Now, using the gauge invariance of the action, we have
\begin{equation}
S\l[\phi_{(s)}\r]=S\l[\phi^{TT}_{(s)}+g\phit_{(s-2)}\r].
\end{equation}
One may further show that
\begin{equation}
S\l[\phi^{TT}+g\phit\r]=S\l[\phi^{TT}\r]+S\l[g\phit\r],
\end{equation}
where 
\begin{equation}
S\l[\phi^{TT}\r]=\phi_{TT}^{\mu_1\ldots\mu_s}\l[\square-\frac{s^2+\l(D-6\r)s-2\l(D-3\r)}{\ell^2}\r]\phi^{TT}_{\mu_1\ldots\mu_s}.
\end{equation}
Therefore, the contribution to the one-loop partition function is given by
\begin{equation}
\cz^{(s)}=\cz^{(s)}_{gh}\l[\det\l(-\square+\frac{s^2+\l(D-6\r)s-2\l(D-3\r)}{\ell^2}\r)_{(s)}\r]^{-\frac{1}{2}}\int\l[D\phit_{(s-2)}\r]e^{-S\l[\phit_{(s-2)}\r]},
\end{equation}
where the subscript $(s)$ on the second term reminds us that the determinant should be evaluated over rank $s$ STT tensors. This expression should be compared to Equation (2.8) of \cite{Gaberdiel:2010ar}. Also, the ghost determinant may be evaluated using the identity
\begin{equation}
1=\int\l[\cd\phi_{\l(s\r)}\r]e^{-\<\phi_{\l(s\r)},\phi_{\l(s\r)}\>}= \cz_{gh}\int\l[\cd\phi_{\l(s\r)}\r]\l[\cd\phit_{\l(s-2\r)}\r]\l[\cd\xi_{\l(s-1\r)}\r]e^{-\<\phi_{\l(s\r)},\phi_{\l(s\r)}\>},
\end{equation}
as in \cite{Gaberdiel:2010xv,Gaberdiel:2010ar}. The evaluation is entirely analogous to the procedure adopted in \cite{Gaberdiel:2010ar}, and we merely mention the final result. The partition function $\cz^{(s)}$ is determined to be a ratio of functional determinants evaluated over STT tensor fields. In particular,
\begin{equation}
\label{determinants}
\ZZ^{(s)}={\l[\det\l(-\square-\frac{\l(s-1\r)\l(3-D-s\r)}{\ell^2}\r)_{(s-1)}\r]^{\frac{1}{2}}\over \l[\det\l(-\square+\frac{s^2+\l(D-6\r)s-2\l(D-3\r)}{\ell^2}\r)_{(s)}\r]^{\frac{1}{2}}}.
\end{equation}
The numerator is a determinant evaluated over rank $s-1$ STT tensor fields. These determinants were evaluated explicitly in \cite{Gopakumar:2011qs} for Laplacians over quotients of $AdS$. In particular, for the thermal quotient of $AdS_{2n+1}$, we find that
\begin{equation}
\log\ZZ^{(s)}=\sum_{m=1}^\infty {e^{-m\beta\l(s+2n-2\r)}\over \l(1-e^{-m\beta}\r)^{2n}} \l(d_s - d_{s-1} e^{-m\beta}\r) ,
\end{equation}
where we adopt the notation $d_s$ for the dimension of the $\l(s,0,\ldots,0\r)$ representation of $SO\l(2n\r)$. We would like to exponentiate this expression to determine the partition function $\ZZ^{(s)}$. The following identity is useful for this purpose.
\begin{equation}
\label{sumidentity}
\sum_{m=1}^\infty  {q^{-m\Delta}\over m\l(1-q^{m}\r)^{2n}} = \sum_{m=1}^\infty {2n+m-2\choose 2n-1} \log {1\over 1-q^{\l(\Delta+m-1\r)}},
\end{equation}
where we have defined $q=e^{-\beta}$. Using \eqref{sumidentity}, we find that
\begin{equation}
\label{singlepart}
\ZZ^{(s)}=\prod_{m=1}^\infty \l({{\l(1-q^{\l(s+m+2n-2\r)}\r)^{d_{s-1} }}\over \l(1-q^{\l(s+m+2n-3\r)}\r)^{d_s}}\r)^{m+2n-2\choose 2n-1}.
\end{equation}
This is the contribution of a single spin-$s$ field to the partition function of the theory in an odd dimensional $AdS$ spacetime. The case of even dimensions is similar, see Appendix \ref{otherdim}. Now, it will be apparent to the reader that many of the subsequent steps that we shall now undertake will follow in arbitrary dimensions as well. However, the resulting expressions are perhaps the nicest for the case of $AdS_5$ and we shall focus on this case from now onwards. We refer the reader to Appendix \ref{otherdim} for the expressions for the blind partition function in $AdS_4$ and $AdS_7$.
\subsection{The Blind Partition Function in $AdS_5$}
\label{blindsection}
While we shall finally work with nonzero chemical potentials for the $SO(4)$ Cartans of $AdS_5$ as well, it is useful, for intuition, to start with the case where they are zero, \textit{i.e.} thermal $AdS_5$. This will also yield some curious forms for the partition function. We have, on setting $n=2$ in \eqref{singlepart},
\begin{equation}
\ZZ^{(s)}=\prod_{m=1}^\infty \l({{\l(1-q^{\l(s+m+2\r)}\r)^{s^2}}\over \l(1-q^{\l(s+m+1\r)}\r)^{(s+1)^2}}\r)^{m+2\choose 3}.
\end{equation}
This is the contribution of a single spin-$s$ field to the partition function of the theory. To determine the full partition function, we need to specify the spectrum of the theory. We choose the spectrum $s=0,1,2,\ldots,\infty$, with each spin appearing only once. The full partition function of the theory is then
\begin{equation}
\ZZ=\prod_{s=0}^\infty \ZZ^{(s)}.
\end{equation}
Note that there are cancellations between the numerator of $\ZZ^{(s)}$ and the denominator of $\ZZ^{(s+1)}$, for $s\geq 1$ because they are both of the form $\l(1-q^{\l(s+m+2\r)}\r)^\a$. We finally obtain
\begin{equation}
\label{blindpartmatch}
\ZZ=\prod_{s=0}^\infty\prod_{m=1}^\infty {1\over\l(1-q^{s+m+2}\r)^{4\l(s+1\r){m+2\choose 3}}}\cdot \prod_{m=1}^\infty {1\over \l(1-q^{m+1}\r)^{{m+2\choose 3}}}.
\end{equation}
We can assemble the first product into a single product over $k\equiv s+n$. This is given by 
\begin{equation}
\prod_{s=0}^\infty\prod_{m=1}^\infty {1\over\l(1-q^{s+m+2}\r)^{4\l(s+1\r){m+2\choose 3}}}=\prod_{k=1}^\infty {1\over \l(1-q^{k+2}\r)^{4{k+4\choose 5}}}.
\end{equation}
We then obtain
\begin{equation}
\label{macmahon}
\ZZ=\prod_{k=1}^\infty {1\over \l(1-q^{k+2}\r)^{4{k+4\choose 5}}}\cdot \prod_{m=1}^\infty {1\over \l(1-q^{m+1}\r)^{{m+2\choose 3}}}.
\end{equation}
This form of the answer is closely related to the $d$-dimensional MacMahon function (see \cite{Balakrishnan:2011bm} for a recent review)
\begin{equation}
\label{dmac}
M_d\l(q\r)=\prod_{n=1}^\infty {1\over \l(1-q^{n}\r)^{{n+d-2\choose d-1}}}.
\end{equation}
We remind the reader that in \cite{Gaberdiel:2010ar} a form for the partition function of a higher-spin theory in $AdS_3$ was obtained in terms of the (two-dimensional) MacMahon function and bore an interpretation as the vacuum character of $\mathcal{W}_{N}$, the asymptotic symmetry algebra of the theory \cite{Henneaux:2010xg, Campoleoni:2010zq}. We also observe the existence of a `nested product' form for this answer. 
\begin{equation}
\label{nested}
\ZZ=\prod_{s_1=3}^\infty \prod_{s_2=s_1}^\infty \prod_{s_3=s_2}^\infty \prod_{s_4=s_3}^\infty \prod_{s_5=s_4}^\infty \prod_{n=s_5}^\infty {1\over\l(1-q^n\r)^4}.
\end{equation}
The reader may verify that this expression does indeed reduce to \eqref{macmahon}. We also remind the reader that an analogous expression for $AdS_3$ was interpreted in \cite{Gaberdiel:2010ar} as the vacuum character of the $\WW_\infty$ algebra. It would be very interesting to see if our five-dimensional answer also admits a similar interpretation.
\subsection{Additional Chemical Potentials in $AdS_5$}
From now onwards we turn on chemical potentials $\l(\beta,\a_1,\a_2\r)$ for $AdS_5$, where $\beta$ is the inverse temperature and $\a_1,\a_2$ are chemical potentials for the $SO\l(4\r)$ Cartans of the $AdS_5$ isometry group $SO(4,2)$. To explicitly evaluate \eqref{determinants}, we will need the character given in equation $5.3$ of \cite{Gopakumar:2011qs}, where we specialise to $n=2$, \textit{i.e.} $AdS_5$, and focus on the case of STT tensors.
\begin{equation}
\label{char1}
\chi_{(\lambda ,s)}\l(\beta,\a_1,\a_2\r)= \frac{e^{-i\beta\lambda}\chi^{SO\l(4\r)}_{\l(s,0\r)}\l(\phi_1,\phi_2\r)+e^{i\beta\lambda}\chi^{SO\l(4\r)}_{\l(s,0\r)}\l(\phi_1,\phi_2\r)}{e^{-2\beta}\vert e^\beta - e^{i\phi_1}\vert^2 \vert e^\beta - e^{i\phi_2}\vert^2}.
\end{equation}
We can evaluate these characters to obtain
\begin{equation}
\chi_{(\lambda ,s)}\l(\beta,\a_1,\a_2\r)= \frac{2\cos\l(\beta\lambda\r)}{e^{-2\beta}\vert e^\beta - e^{i\phi_1}\vert^2 \vert e^\beta - e^{i\phi_2}\vert^2}{\sin\l(\l(s+1\r)\a_1\r)\over \sin\l(\a_1\r)}{\sin\l(\l(s+1\r)\a_2\r)\over \sin\l(\a_2\r)},
\end{equation}
where the $\phi$s and $\a$s are related by
\begin{equation}
\phi_1=\a_1+\a_2, \quad \phi_2=\a_1-\a_2.
\end{equation}
The partition function of a massless spin-$s$ particle may then be computed to obtain
\begin{equation}
\label{refinedpart}
\log\ZZ^{(s)} = \sum_{m=1}^\infty {1\over m}{e^{-m\beta \l(s+2\r)}\over \vert 1-e^{-m\l(\beta-i\phi_1\r)}\vert ^2 \vert 1-e^{-m\l(\beta-i\phi_2\r)}\vert ^2} \l[\chi^{SO\l(4\r)}_{\l(s,0\r)}-\chi^{SO\l(4\r)}_{\l(s-1,0\r)}e^{-m\beta}\r],
\end{equation}
where the $SO(4)$ characters are evaluated over the angles $\l(m\phi_1,m\phi_2\r)$. This is precisely the answer that would be obtained via a Hamiltonian computation, as carried out in \cite{Gibbons:2006ij} using the results of \cite{Dolan:2005wy,Barabanschikov:2005ri}.

\section{The Refined Partition Function in $AdS_5$}
\label{refinedsection}
Having acquired some intuition for the kind of simplifications we may expect by looking at the blind partition function, we shall now now turn to the case of non-zero chemical potentials $\a_1$ and $\a_2$ and exponentiate the expression \eqref{refinedpart}. This will enable us to write down another nested form for the partition function. It would be again useful to introduce notation
\begin{equation}
e^{-\beta}=q,\quad e^{i\a_1}=p,\quad e^{i\a_2}=r.
\end{equation}
We finally obtain
\begin{equation}
\ZZ^{(s)}=\prod_{m_i=0}^\infty {\prod_{k,l=0}^{s-1} \l(1- q^{m_1+m_2+m_3+m_4+s+3}p^{m_1+m_3+s-1}r^{m_1+m_4+s-1}\pb^{m_2+m_4+2k}\rb^{m_2+m_3+2l}\r)\over \prod_{k,l=0}^{s} \l(1- q^{m_1+m_2+m_3+m_4+s+2}p^{m_1+m_3+s}r^{m_1+m_4+s}\pb^{m_2+m_4+2k}\rb^{m_2+m_3+2l}\r)},
\end{equation}
where the product over $m_i$ collectively denotes a product over $m_1,\ldots,m_4$. It turns out that even for the refined case, the ratio of the numerator of $\ZZ^{(s)}$ and the denominator of $\ZZ^{(s+1)}$ is again simple. This leads to the following form for the one-loop partition function of the theory.
\begin{equation}
\label{unsimpleZ}
\ZZ=\prod_{s=0}^\infty\prod_{m_i=0}{1\over \hat{\prod} \l(1- q^{m_1+m_2+m_3+m_4+s+3}p^{m_1+m_3+s-1}r^{m_1+m_4+s-1}\pb^{m_2+m_4+2k}\rb^{m_2+m_3+2l}\r) },
\end{equation}
where the hatted product in the denominator runs over the following values of $k,l$, 
\begin{equation*}\begin{split}
k=0,\,l=0;\quad k=0,\,l=s+1;\quad k=s+1,\,l=0 \quad k=s+1,\,l=s+1\\ k=0,\,l=1\ldots s;\quad l=0,\,k=1\ldots s;\quad k=s+1,\,l=1\ldots s;\quad l=s+1,\,k=1\ldots s,\end{split}
\end{equation*}
and we have used the fact that $(p\pb)^2(r\rb)^2=1$. 
 It is useful to define the following combinations
\begin{equation}
q_1=pr,\quad q_2= \bar{p}\bar{r},\quad q_3=p\bar{r},\quad q_4=\bar{p}r.
\end{equation}
The expression \eqref{unsimpleZ} may be simplified using the procedure outlined in Appendix \ref{simplifications}. We finally find that the partition function may be written in terms of the product of four nested product expressions.
\begin{equation}
\label{nestedZ}
\ZZ\equiv \ZZ_{(0)}\cdot \ZZ_{(1)}\cdot\ZZ_{(2)}\cdot\ZZ_{(3)}\cdot\ZZ_{(4)},
\end{equation}
where $\ZZ_{(0)}$ is the partition function of the scalar, which may be computed to obtain
\begin{equation}
\ZZ_{(0)}= \prod_{m_{1,2,3,4}} {1\over \l(1-q^2\l(qq_1\r)^{m_1}\l(qq_2\r)^{m_2}\l(qq_3\r)^{m_3}\l(qq_4\r)^{m_4}\r)},
\end{equation}
and 
\begin{equation}
\ZZ_{(1)}=\prod_{k=0}^\infty \prod_{m_{2,3,4}}^\infty \prod_{s=1}^\infty \prod_{n=s}^\infty {1\over\l(1-\l(qq_1\r)^n q^{k+m_2+m_3+m_4+2}q_1^k q_2^{m_2+k}q_3^{m_3+k}q_4^{m_4}\r)}.
\end{equation}
This has been explicitly evaluated in appendix \ref{simplifications}. The other $\ZZ_{(i)}$s have similar expressions which we now enumerate.
\begin{equation}
\label{nested2}
\ZZ_{(2)}=\prod_{k=0}^\infty \prod_{m_{1,2,4}}^\infty \prod_{s=1}^\infty \prod_{n=s}^\infty {1\over\l(1-\l(qq_3\r)^n q^{k+m_1+m_2+m_4+2}q_1^{m_1} q_2^{m_2+k}q_3^{k}q_4^{m_4+k}\r)},
\end{equation}
\begin{equation}
\label{nested2}
\ZZ_{(3)}=\prod_{k=0}^\infty \prod_{m_{1,2,3}}^\infty \prod_{s=1}^\infty \prod_{n=s}^\infty {1\over\l(1-\l(qq_4\r)^n q^{k+m_1+m_2+m_4+2}q_1^{m_1+k} q_2^{m_2}q_3^{m_3+k}q_4^{k}\r)},
\end{equation}
\begin{equation}
\label{nested2}
\ZZ_{(4)}=\prod_{k=0}^\infty \prod_{m_{1,3,4}}^\infty \prod_{s=1}^\infty \prod_{n=s}^\infty {1\over\l(1-\l(qq_2\r)^n q^{k+m_1+m_3+m_4+2}q_1^{m_1+k} q_2^{k}q_3^{m_3}q_4^{m_4+k}\r)}.
\end{equation}
\section{Discussion}
\label{conclusions}
In this paper, we applied the heat kernel results of \cite{Gopakumar:2011qs} to compute the partition function of a higher-spin theory in $AdS_5$. To get a better feeling for the possible content of the answer it is useful to recollect elements of the corresponding story for pure gravity and higher-spin gravity in $AdS_3$. This discussion is qualitative, and we refer the reader to the original papers for more details. As we remarked previously, the asymptotic symmetry algebra of pure gravity on $AdS_3$ comprises of two copies of the Virasoro algebra \cite{BH} and for higher-spin gravity comprises of two copies of the $\WW$ algebra \cite{Henneaux:2010xg,Campoleoni:2010zq}, while the gauge symmetry is $SL(2,R)\times SL(2,R)$ and $SL(N,R)\times SL(N,R)$ or $hs(1,1)$ respectively. One manifestation of the asymptotic symmetry algebra is in the one-loop partition function of the theories expanded about their $AdS$ vacua. For the higher-spin theory with spectrum of spins $s=2,\ldots,\infty$, the partition function is \cite{Gaberdiel:2010ar}
\begin{equation}
\mathcal{Z}=\prod_{s=2}^{\infty}\prod_{n=s}^{\infty}\frac{1}{\vert 1-q^n\vert ^2}= \chi_0\l(\WW_\infty\r)\times \bar{\chi}_0\l(\WW_\infty\r).
\end{equation}
This is (two copies of) the character evaluated over the Verma module built out of the $AdS_3$ vacuum as the lowest weight state and the raising operators $\mathcal{W}^{(s)}_{-m}$, $s=2,\ldots,\infty$, acting on it. This is the sense in which the second-quantised partition function is supposed to encode information about multiparticle symmetries of the theory. The corresponding expressions for pure gravity \cite{Maloney:2007ud,Giombi:2008vd,David:2009xg} are perhaps more familiar to the reader. The partition function is then a character of the Virasoro algebra, again with the $AdS_3$ vacuum as the lowest weight state and the generators $L_{-m}$ acting as raising operators, for $m\geq 2$. These generators, when acting on the $AdS_3$ vacuum generate large gauge transformations to create multiparticle states---boundary gravitons---over the $AdS_3$ vacuum.

We obtained a form for the answer that is suggestive of a vacuum character of a larger symmetry group than the higher-spin symmetry. In particular, following the reasoning suggested above, it leads us to speculate that there are additional generators in the symmetry algebra that act on multiparticle states built out of twist-two operators, which sit in the vacuum representation of the symmetry algebra. It would be very interesting to verify if this is indeed the case. A natural candidate symmetry to relate these results to is the Yangian symmetry of $\mathcal{N}=4$ SYM. However, we do not know of an apparent connection, especially since the Yangian algebra does not close over the set of gauge-invariant operators.\footnote{We thank N. Beisert for this remark.} Alternatively, this might be related to the additional symmetries of free Yang-Mills noted in \cite{Kimura:2008ac}. There is additionally the intriguing possibility that these might be connected to multiparticle symmetries already noted in the literature for higher-spin theories \cite{Vasiliev:2001zy,Gelfond:2003vh,Gelfond:2010pm}.\footnote{We thank M. Vasiliev for this observation.} 

\textit{Note:} After an initial version of this paper appeared on the arXiv, we learned of \cite{Bandos:1999qf,Plyushchay:2003gv,Bandos:2004nn}, in which additional enhanced symmetries for higher-spin theories had been observed.\footnote{We thank D. Sorokin for bringing this to our attention.} It is also possible that our results are related to these symmetries.

We leave the exploration of these questions to future work.
\section*{Acknowledgements}
We would like to thank Massimo Bianchi, Niklas Beisert, Dileep Jatkar, Matthias Gaberdiel, Suresh Govindarajan, Jose Morales, Albrecht Klemm, Suvrat Raju, Arunabha Saha, Ashoke Sen, Xi Yin, and Mikhail Vasiliev for very helpful discussions. We would especially like to thank Rajesh Gopakumar for initial collaboration and several very helpful discussions, as well as his comments on the manuscript. SL would like to thank CERN, ETH Zurich, INFN Rome, Tor Vergata, SNS Pisa, TU Vienna, UvA Amsterdam, ULB Brussels, Utrecht University and the organisers and participants of NSM 2011 and the $6^{th}$ Asian Winter School on Strings, Particles and Cosmology for the opportunity to present an initial version of these results, and very helpful comments. SL would also like to thank CHEP (IISc Bangalore) and ICTS Bangalore for hospitality while some of this work was carried out, and more generally the people of India for their support to research in basic science. The work of RG is partially supported by the ERC Advanced Grant no. 246974, ``Supersymmetry: a window to non-perturbative physics''.

\appendix
\section{Blind Partition functions for $AdS_4$ and $AdS_7$}
\label{otherdim}
We shall begin by using \eqref{singlepart} to obtain the blind partition function of the Vasiliev theory in $AdS_7$. Observations parallel to those in Section \ref{blindsection} lead us to conclude that the expressions for the partition function are again simple. We need to evaluate the product
\begin{equation}
\prod_{s=1}^\infty \prod_{m=1}^\infty {1\over \l(1-q^{s+m+4}\r)^{{m+4\choose 5}\l(d_{s+1}-d_{s-1}\r)}},
\end{equation}
where $d_s$ denotes the $\l(s,0,0\r)$ representation of $SO\l(6\r)$. Computing these dimensions (see \cite{Gopakumar:2011qs} and references therein), we find that the partition function of the Vasiliev theory may be expressed as the product
\begin{equation}
\ZZ=\ZZ_{(0)}\cdot\prod_{s=0}^\infty\prod_{m=1}^\infty{1\over \l(1-q^{s+m+4}\r)^{{m+4\choose 5}\l(d_{s+1}-d_{s-1}\r)}},
\end{equation}
where $\ZZ_{(0)}$ is the partition function of the scalar, given by
\begin{equation}
\ZZ_{(0)}=\prod_{m=1}^\infty {1\over\l(1-q^{m+3}\r)^{m+4\choose 5}}.
\end{equation}
We finally obtain
\begin{equation}
\ZZ=\prod_{m=1}^\infty {1\over\l(1-q^{m+3}\r)^{m+4\choose 5}}\cdot {1\over\l(1-q^{m+4}\r)^{4{m+8\choose 9}+{m+6\choose 7}+{m+5\choose 6}}},
\end{equation}
which is again related to the $d$-dimensional MacMahon function \eqref{dmac}.

We now turn to the case of $AdS_4$. The entire heat kernel analysis of this paper and \cite{Gopakumar:2011qs} may be applied to even dimensional hyperboloids. We merely mention the final result for $AdS_4$. We find that
\begin{equation}
\ZZ={1\over\l(1-q\r)\l(1-q^2\r)^3}\prod_{m=2}^\infty {1\over \l(1-q^m\r)^{m+1\choose 2}\l(1-q^{m+1}\r)^{3{m+1\choose 2}+4 {m+2\choose 3}}}.
\end{equation}
This is again related to the $d$-dimensional MacMahon function \eqref{dmac}.

\section{Computing the Refined Partition Function}
\label{simplifications}
We outline some of the main steps involved in reducing \eqref{unsimpleZ} to the nested form \eqref{nestedZ}. We begin by expanding out $\hat{\prod}$ that appears in \eqref{unsimpleZ} into two products $A$ and $B$. We will evaluate
\begin{equation}
\begin{split}
A = &\prod_{m_i=0}^\infty \prod_{s=0}^\infty \l(1-q^{m_1+m_2+m_3+m_4+s+3}p^{m_1+m_3+s+1}r^{m_1+m_4+s+1}\bar{p}^{m_2+m_4}\bar{r}^{m_2+m_3}\r) \\
&\prod_{m_i=0}^\infty \prod_{s=0}^\infty \l(1-q^{m_1+m_2+m_3+m_4+s+3}p^{m_1+m_3+s+1}r^{m_1+m_4+s+1}\bar{p}^{m_2+m_4}\bar{r}^{m_2+m_3+2(s+1)}\r) \\ &\prod_{m_i=0}^\infty \prod_{s=0}^\infty \l(1-q^{m_1+m_2+m_3+m_4+s+3}p^{m_1+m_3+s+1}r^{m_1+m_4+s+1}\bar{p}^{m_2+m_4+2(s+1)}\bar{r}^{m_2+m_3}\r) \\ &\prod_{m_i=0}^\infty \prod_{s=0}^\infty \l(1-q^{m_1+m_2+m_3+m_4+s+3} p^{m_1+m_3+s+1} r^{m_1+m_4+s+1}\bar{p}^{m_2+m_4+2(s+1)}\bar{r}^{m_2+m_3+2(s+1)}\r),
\end{split}
\end{equation}
and 
\begin{equation}
\begin{split}
B = &\prod_{m_i=0}^\infty \prod_{s=1}^\infty \prod_{k=1}^s\l(1-q^{m_1+m_2+m_3+m_4+s+3}p^{m_1+m_3+s+1}r^{m_1+m_4+s+1}\bar{p}^{m_2+m_4}\bar{r}^{m_2+m_3+2k}\r) \\ &\prod_{m_i=0}^\infty \prod_{s=1}^\infty \prod_{k=1}^s\l(1-q^{m_1+m_2+m_3+m_4+s+3}p^{m_1+m_3+s+1}r^{m_1+m_4+s+1}\bar{p}^{m_2+m_4+2k}\bar{r}^{m_2+m_3+2(s+1)}\r) \\ &\prod_{m_i=0}^\infty \prod_{s=1}^\infty \prod_{k=1}^s \l(1-q^{m_1+m_2+m_3+m_4+s+3}p^{m_1+m_3+s+1}r^{m_1+m_4+s+1}\bar{p}^{m_2+m_4+2k} \bar{r}^{m_2+m_3}\r) \\ &\prod_{m_i=0}^\infty \prod_{s=1}^\infty \prod_{k=1}^s \l(1-q^{m_1+m_2+m_3+m_4+s+3}p^{m_1+m_3+s+1}r^{m_1+m_4+s+1}\bar{p}^{m_2+m_4+2(s+1)}\bar{r}^{m_2+m_3+2k}\r).
\end{split}
\end{equation}
The full partition function of the theory is the inverse of the product of $A$ and $B$. We will now simplify these expressions.

\begin{equation}
\begin{split}
A = &\prod_{m_i,s}\l(1-q^{m_1+m_2+m_3+m_4+s+3}q_1^{m_1+s+1}q_2^{m_2}q_3^{m_3}q_4^{m_4}\r) \l(1-q^{m_1+m_2+m_3+m_4+s+3}q_1^{m_1} q_2^{m_2} q_3^{m_3+s+1}q_4^{m_4}\r)\\ &\prod_{m_i,s}\l(1-q^{m_1+m_2+m_3+m_4+s+3}q_1^{m_1} q_2^{m_2} q_3^{m_3} q_4^{m_4+s+1}\r) \l(1-q^{m_1+m_2+m_3+m_4+s+3}q_1^{m_1} q_2^{m_2+s+1} q_3^{m_3} q_4^{m_4}\r),
\end{split}
\end{equation}
where we have used the fact that $p\bar{p}=1=r\bar{r}$. In each product, there is a pairing of one of $m_i$ and $s$. We use this to write 
\begin{equation}
\begin{split}
A=&\prod_{m_i=0}^\infty \l(1-q^{m+s+3}q_1^{m_1+1}q_2^{m_2}q_3^{m_3}q_4^{m_4}\r)^{m_1+1}\prod_{m_i=0}^\infty \l(1-q^{m+s+3}q_1^{m_1} q_2^{m_2} q_3^{m_3+1}q_4^{m_4}\r)^{m_3+1} \\ &\prod_{m_i=0}^\infty \l(1-q^{m+s+3}q_1^{m_1} q_2^{m_2} q_3^{m_3} q_4^{m_4+1}\r)^{m_4+1} \prod_{m_i=0}^\infty\l(1-q^{m+s+3}q_1^{m_1} q_2^{m_2+1} q_3^{m_3} q_4^{m_4}\r)^{m_2+1},
\end{split}
\end{equation}
where we have defined $m \equiv \sum_i m_i$.
Now each product is of the form
\begin{equation}
\prod_{n=1}^\infty\l(1-\alpha q^n\r)^n= \prod_{s=1}^\infty \prod_{n=s}^\infty \l(1-\alpha q^n\r),
\end{equation}
where $\alpha$ is independent of $n$. This gives a (two-dimensional) MacMahon function form for this sector of the partition function, which we can arrange into a nested form. For example, the first product reduces to
\begin{equation}
\label{m0eqn}
\prod_{m_2,m_3,m_4=0}^\infty \l(\prod_{s=1}^\infty \prod_{n=s}^\infty \l(1-\l(qq_1\r)^n q^{m_2+m_3+d+2}q_2^{m_2}q_3^{m_3}q_4^{m_4}\r)\r).
\end{equation}
There are similar expressions for the other three products, which arise by permuting $q_i$'s above.
We will now look at the $B$ term, and obtain its nested product representation. Written in terms of the $q_i$s, the first product that enters in $B$ is 
\begin{equation}
\begin{split}
\prod_{m_i=0}^\infty \prod_{s=1}^\infty &\prod_{k=1}^s \l(1-\l(qq_1\r)^{(m_1+s+1)}q^{m_2+m_3+d+2}q_2^{m_2+k}q_3^{m_3+k}q_4^{m_4}\r)\\ &= \prod_{m_i=0}^\infty \prod_{k=1}^\infty \prod_{s=k}^\infty \l(1-\l(qq_1\r)^{(m_1+s+1)}q^{m_2+m_3+d+2}q_2^{m_2+k}q_3^{m_3+k}q_4^{m_4}\r),
\end{split}
\end{equation}
where we have changed the products over $k$ and $s$. This enumerates the same combinations $(k,s)$ as the previous product. Again, redefining $a+s$ to $a$, and $s-k$ to $s$, we get
\begin{equation}
\begin{split}
\prod_{m_i=0}^\infty \prod_{k=1}^\infty &\prod_{s=0}^s \l(1-\l(qq_1\r)^{(m_1+s+1)}q^{k+m_2+m_3+m_4+2}q_1^k q_2^{m_2+k}q_3^{m_3+k}q_4^{m_4}\r)\\ &= \prod_{k=1}^\infty \prod_{m_i=0}^\infty   \l(1-\l(qq_1\r)^{(m_1+1)}q^{k+m_2+m_3+m_4+2}q_1^k q_2^{m_2+k}q_3^{m_3+k}q_4^{m_4}\r)^{m_1+1},
\end{split}
\end{equation}
from which we can write the nested form
\begin{equation}
\prod_{k=1}^\infty \prod_{m_{2,3,4}}^\infty \prod_{s=1}^\infty \prod_{n=s}^\infty \l(1-\l(qq_1\r)^n q^{k+m_2+m_3+m_4+2}q_1^k q_2^{m_2+k}q_3^{m_3+k}q_4^{m_4}\r)
\end{equation}
Note that \eqref{m0eqn} is just the $k=0$ term in this product. We can write the contribution of these two terms together as
\begin{equation}
\label{nested1}
\ZZ_{(1)}^{-1}=\prod_{k=0}^\infty \prod_{m_{2,3,4}}^\infty \l(\prod_{s=1}^\infty \prod_{n=s}^\infty \l(1-\l(qq_1\r)^n q^{k+m_2+m_3+m_4+2}q_1^k q_2^{m_2+k}q_3^{m_3+k}q_4^{m_4}\r)\r).
\end{equation}
The other three contributions to the partition function may be similarly written down by combining the second, third and fourth terms respectively from $A$ and $B$, we obtain the nested product forms
\begin{equation}
\ZZ_{(2)}^{-1}=\prod_{k=0}^\infty \prod_{m_{1,2,4}}^\infty \l(\prod_{s=1}^\infty \prod_{n=s}^\infty \l(1-\l(qq_3\r)^n q^{k+m_1+m_2+m_4+2}q_1^{m_1} q_2^{m_2+k}q_3^{k}q_4^{m_4+k}\r)\r),
\end{equation}
\begin{equation}
\ZZ_{(3)}^{-1}=\prod_{k=0}^\infty \prod_{m_{1,2,3}}^\infty \l(\prod_{s=1}^\infty \prod_{n=s}^\infty \l(1-\l(qq_4\r)^n q^{k+m_1+m_2+m_4+2}q_1^{m_1+k} q_2^{m_2}q_3^{m_3+k}q_4^{k}\r)\r),
\end{equation}
\begin{equation}
\ZZ_{(4)}^{-1}=\prod_{k=0}^\infty \prod_{m_{1,3,4}}^\infty \l(\prod_{s=1}^\infty \prod_{n=s}^\infty \l(1-\l(qq_2\r)^n q^{k+m_1+m_3+m_4+2}q_1^{m_1+k} q_2^{k}q_3^{m_3}q_4^{m_4+k}\r)\r).
\end{equation}
Putting these expressions together yields the expression for $\ZZ$ in \eqref{nestedZ}.

\section{A Consistency Check} 
We finally show that the expressions obtained in section \ref{refinedsection} for the refined partition function \eqref{nestedZ} are consistent with the blind partition function. To do so, we shall set all the $q_i$s to one. Then \eqref{nestedZ} reduces to
\begin{equation} 
\label{anynested} 
\prod_{mabc=0}^\infty \prod_{s=1}^\infty \prod_{n=s}^\infty\l(1-q^nq^{m+a+b+c+2}\r)^4.
\end{equation} 
We arrange this into a sum over $d=m+a+b+c$. We obtain
\begin{equation}
\prod_{d=0}^\infty \prod_{s=0}^\infty \prod_{n=0}^\infty \l(1-q^{d+3}q^{n+s}\r)^{4\binom{d+3}{3}}.
\end{equation}
Then, the sum over $n$ and $s$ can be converted to a sum over $s$. $d$ can be renamed to $n$, and shifted by $1$, and we finally obtain
\begin{equation}
\prod_{s=0}^\infty\prod_{n=1}^\infty \l(1-q^{n+s+2}\r)^{4(s+1)\binom{n+2}{3}}.
\end{equation}
This precisely matches with the first product that appears in \eqref{blindpartmatch}.

We can also obtain the nested form \eqref{nested} from the expression \eqref{anynested}. to do so, we start with \eqref{anynested}
\begin{equation}
\prod_{mabc=0}^\infty \prod_{s=1}^\infty \prod_{n=s}^\infty\l(1-q^nq^{m+a+b+c+2}\r)^4=  \prod_{s=3}^\infty \prod_{n=s}^\infty\prod_{m=0}^\infty\prod_{a=0}^\infty\prod_{b=0}^\infty\prod_{c=0}^\infty\l(1-q^nq^{m+a+b+c}\r)^4.
\end{equation}
Now on defining $n^\prime=n+m$, we may write
\begin{equation}
\prod_{n=s}^\infty\prod_{m=0}^\infty \l(1-q^{n+m}q^{a+b+c}\r)^4=\prod_{s_2=s}^\infty\prod_{n^\prime=s_2}^\infty \l(1-q^{n^\prime+a+b+c}\r)^4, 
\end{equation}
and hence \eqref{anynested} reduces to 
\begin{equation}
\prod_{s=3}^\infty\prod_{s_2=s}^\infty\prod_{n^\prime=s_2}^\infty\prod_{a=0}^\infty\prod_{b=0}^\infty\prod_{c=0}^\infty\l(1-q^{n^\prime+a+b+c}\r)^4.
\end{equation}
We can similarly absorb the products over $a,b,c$ to write \eqref{anynested} as
\begin{equation}
\prod_{s=3}^\infty \prod_{s_2=s}^\infty \prod_{s_3=s_2}^\infty \prod_{s_4=s_3}^\infty \prod_{s_5=s_4}^\infty \prod_{N=s_5}^\infty \l(1-q^N\r)^4,
\end{equation}
which is the familiar form that appears in \eqref{nested} with the replacements $s\rightarrow s_1$, and $N\rightarrow n$.

\end{document}